\documentclass{PoS}

\usepackage{amsmath}
\usepackage{amssymb}
\usepackage{epsfig}

\title{BQCD -- Berlin quantum chromodynamics program}

\ShortTitle{BQCD}

\author{Yoshifumi Nakamura\\
Institut für Theoretische Physik, Universität Regensburg, 
93040 Regensburg, Germany \\
Center for Computational Sciences, University of Tsukuba, 
Tsukuba, Ibaraki 305-8577, Japan \\
E-mail: \email{yoshi@ccs.tsukuba.ac.jp}}

\author{\speaker{Hinnerk St\"uben}\\
Konrad-Zuse-Zentrum f\"ur Informationstechnik Berlin, 14195 Berlin, Germany\\
E-mail: \email{stueben@zib.de}}

\abstract{We publish BQCD as free software under the GNU General Public
  License.  BQCD is a Hybrid Monte-Carlo program that simulates lattice
  QCD with dynamical Wilson fermions.  It is one of the main production
  programs of the QCDSF collaboration.  The program can simulate $2$ and
  $2 + 1$ fermion flavours with pure, clover improved, and stout smeared
  fat link Wilson fermions as well as standard plaquette, and an
  improved (rectangle) gauge action.  The single flavour is simulated
  with the Rational Hybrid Monte-Carlo algorithm.}

\FullConference{The XXVIII International Symposium on Lattice Field Theory, Lattice2010\\
		June 14-19, 2010\\
		Villasimius, Italy}

\newcommand{\Wilson}{\mathrm{Wilson}}
\newcommand{\Det}{\mathrm{det}}        % \det is already defined
\newcommand{\csw}{c_{\mathrm SW}}
\renewcommand{\Re}{\mathop{\mathrm{Re}}}
\newcommand{\Tr}{\mathop{\mathrm{Tr}}}

\newlength{\dWidth}     % digit width
\newcommand{\D}{\makebox[\dWidth]{}}

\begin{document}
%==============================================================================
\section{Introduction}

Berlin quantum chromodynamics program (BQCD) is a Hybrid Monte-Carlo
\cite{Duane:1987de} program for simulating lattice QCD with dynamical
Wilson fermions.  The development of BQCD started in 1998 for the two
flavour case and the standard Wilson action.  It was written for a study
of parallel tempering \cite{Ilgenfritz:2001jp}.  At that time the whole
parallelisation framework was completed.  Soon the program was extended
in two different directions.  The first direction was the implementation
of clover $O(a)$ improvement of the fermion action.  With the
availability of clover improvement BQCD became one of the main
production codes of the QCDSF collaboration \cite{Stuben:2000wm}.  The
second direction was the addition of an external field to the standard
Wilson action in order to study the Aoki phase \cite{Ilgenfritz:2003gw}.
The next milestone was the implementation of the Hasenbusch trick
\cite{Hasenbusch:2001ne,AliKhan:2003mu}.  Starting in 2006 the code has
been largely extended to enable simulations including a third fermion
flavour
\cite{Gockeler:2007rm,Cundy:2008mb,Cundy:2009yy,Bietenholz:2009fi}.
This extension includes the implementation of Rational Hybrid
Monte-Carlo (RHMC) \cite{Clark:2003na} for the simulation of the third
quark flavour as well as many algorithmic and performance improvements.

The code is also being used by the DIK Collaboration for simulations at
finite temperature \cite{Nakamura:2004re,Bornyakov:2009qh}.  Several
people took BQCD as a starting point for adding their own code for
measurements.  The plan of the QPACE project \cite{Baier:2009yq} to port
BQCD, in particular the fermion matrix multiplication and solvers 
\cite{andrea-lat10}, to their machine 
has triggered the publication of the code as free
software under the \textsl{GNU General Public License} on the occasion
of this Lattice conference.  The source and a manual can be downloaded
from \cite{bqcd}.  A description on building and testing binaries can be
found in the manual.

%==============================================================================
\section{Actions}

The program can simulate the QCD with the following actions. 
The gauge action can be the Wilson action 

\begin{equation}
S = S^\Wilson_G =
\sum_{\mathrm{plaquette}} 
\frac{1}{3}\, \Re\,\Tr\,(1 - U_{\mathrm{plaquette}}) 
\end{equation}
or a Symanzik improved gauge action
\begin{equation}
S_G =
\frac{6}{g^2}\left[ 
   c_0 \sum_{\rm plaquette}\frac{1}{3}\, \Re\,\Tr\, (1-U_{\rm plaquette}) 
 + c_1 \sum_{\rm rectangle}\frac{1}{3}\, \Re\,\Tr\, (1-U_{\rm rectangle}) 
             \right] \,,
\end{equation}
with $c_0 + 8 c_1  = 1$.  The fermion action can be the Wilson action
\begin{equation}
S^\Wilson_F = \sum_x
\left\{
\bar{\psi}(x) \psi(x) 
- \kappa \left[
\bar{\psi}(x) U_\mu^\dagger(x-\hat{\mu})(1+\gamma_\mu) \psi(x-\hat{\mu})
+\bar{\psi}(x) U_\mu(x)(1-\gamma_\mu) \psi(x+\hat{\mu})
\right]
\right\}\,,
\end{equation}
the Wilson action plus an explicitly parity-flavour symmetry breaking
source term, where $\tau^3$ is the third Pauli matrix
\begin{equation}
S_F = S^\Wilson_F + h \sum_x \bar{\psi}(x) i \gamma_5 \tau^3 \psi(x)\,,
\end{equation}
the clover action
\begin{equation}
S_F = S^\Wilson_F - \frac{i}{2} \kappa\,\csw 
\sum_x \bar{\psi}(x) \sigma_{\mu\nu} F_{\mu\nu}(x) \psi(x)\,,
\end{equation}
the clover action plus a CP breaking term
\begin{equation}
S_F = S^\Wilson_F - \frac{i}{2} \kappa\,\csw 
\sum_x \bar{\psi}(x) \sigma_{\mu\nu} F_{\mu\nu}(x) \psi(x) + \theta \bar{\psi}(x) \gamma_5 \psi(x)
\end{equation}
or a stout smeared fat link action (any term containing gauge links can
be smeared), in particular the SLiNC fermion action \cite{Cundy:2009yy}
\begin{equation}
\begin{split}
S_F = \sum_x \Big\{\bar{\psi}(x)\psi(x)
&- \kappa\, \bar{\psi}(x) U_\mu^\dagger(x-\hat{\mu})[1+\gamma_\mu]
\psi(x-\hat{\mu})\\[0.0em]
&- \kappa\, \bar{\psi}(x) U_\mu(x)[1-\gamma_\mu]
\psi(x+\hat{\mu})
+\frac{i}{2} \kappa \, \csw\, \bar{\psi}(x)
\sigma_{\mu\nu} F_{\mu\nu}(x) \psi(x) \Big\} \, ,
\end{split}
\end{equation}
where the gauge links $U_\mu$ are replaced by stout links~\cite{Morningstar:2003gk}
\begin{equation}
U_\mu \rightarrow \tilde{U}_\mu(x) = e^{iQ_\mu(x)} \, U_\mu(x)\,,
\end{equation}
with
\begin{equation}
Q_\mu(x)=\frac{\alpha}{2i} \left[V_\mu(x) U_\mu^\dagger(x) -
  U_\mu(x)V_\mu^\dagger(x) -\frac{1}{3} {\rm Tr} \,\left(V_\mu(x)
  U_\mu^\dagger(x) -  U_\mu(x)V_\mu^\dagger(x)\right)\right] \,,
\end{equation}
where $V_\mu(x)$ is the sum over all staples associated with the link.
Boundary condition for the gauge field are periodic in all directions.  For
the fermions boundary conditions can be chosen to be anti-periodic or periodic
for each dimension.

%==============================================================================
\section{Observables}

The following gluonic observables can be measured: the average plaquette and
average rectangular plaquette, the topological charge (the topological charge
is measured with the field theoretic method after cooling the gauge field
configuration), the Polyakov loop.  
In addition some fermionic bulk quantities can be measured (from stochastic
estimators):

\begin{eqnarray*}
\langle\bar{\psi}\psi\rangle & = & \frac{1}{12V}\langle\Tr(M^{-1})\rangle 
\qquad \mbox{('chiral condensate')} \\
\langle\bar{\psi}\gamma_5\psi\rangle & = & \frac{1}{12V}\langle\Tr(\gamma_5M^{-1})\rangle \\
\langle\Pi^2\rangle & = & \frac{1}{12V}\langle\Tr(M^{\dagger}M)^{-1}\rangle
\qquad \mbox{('pion norm')} 
\end{eqnarray*}

%==============================================================================
\section{Algorithmic improvements}

\subsection{Integrators}

HMC trajectories can be integrated with leapfrog or Omelyan
\cite{Takaishi:2005tz} integrators.  Multi timescale integration is
possible with up to six time scales.  In the following we explain a
multi timescale setup that is used in production for $N_f = 2 + 1$
improved Wilson fermions.  Starting point is the partition function
\begin{eqnarray}
Z &=& \int DU D\bar{\psi} D\psi e^{-S} \\
S &=& S_g(\beta) + S_{l}(\kappa_{l},c_{\rm SW}) + S_s(\kappa_{s},c_{\rm SW})
\end{eqnarray}
where $S_g$ is a gluon action, $S_l$ is an action for the degenerate
$u$- and $d$-quarks and $S_s$ is an action for the strange quark. After
integrating out fermions
\begin{equation}
  S=S_g(\beta) - \ln  [\det M_{l}^{{\dagger}}M_{l}] 
                      [\det M_{s}^{{\dagger}}M_{s}]^{1 \over 2}\,.
\end{equation}
First even-odd preconditioning is applied
\begin{eqnarray}
\det M_{l}^{\dagger} M_{l}
&\propto& \det(1+T_{oo}^{l})^2 \det Q_{l}^{\dagger}Q_{l} \\
~[\det M_{s}^{\dagger} M_{s}]^{1 \over 2}
&\propto& \det(1+T_{oo}^{s}) [\det Q_{s}^{\dagger}Q_{s}]^{1 \over 2}
\end{eqnarray}
where 
\begin{eqnarray}
 Q &=& (1+T)_{\rm ee} - M_{\rm eo} (1+T)^{-1}_{\rm oo} M_{\rm oe} \\
 T &=& \frac{\rm i}{2} c_{\rm SW}\, \kappa\, \sigma_{\mu\nu} F_{\mu\nu}\,.
\end{eqnarray}
Then $\det Q_{l}^{\dagger}Q_{l}$ is separated following
Hasenbusch~\cite{Hasenbusch:2001ne}
\begin{equation}
\det Q_{l}^{\dagger}Q_{l}
= \det W_{l}^{\dagger} W_{l} 
\det {Q_{l}^{\dagger}Q_{l} \over W_{l} W_{l}^{\dagger} }\,,
\qquad W=Q+\rho\,.
\end{equation}
Finally the standard action is modified to
\begin{equation}
S=S_g + S_{\Det}^{l} + S_{\Det}^{s} + S_{f1}^{l} + S_{f2}^{l} + S_{fr}^{s} \,,
\end{equation}
where
\begin{eqnarray}
S_{\Det}^{l} &=& -2\, \Tr\, \log[1+T_{\rm oo}(\kappa^l)] \\
S_{\Det}^{s} &=& - \, \Tr\, \log[1+T_{\rm oo}(\kappa^s)] \\
S_{f1}^{l}  &=& \phi_1^{\dagger} [W(\kappa^l)^{\dagger}W(\kappa^l)]^{-1}  \phi_1 \\
S_{f2}^{l}  &=& \phi_2^{\dagger}W(\kappa^l)[Q(\kappa^l)^{\dagger}Q(\kappa^l)]^{-1}
                                        W(\kappa^l)^{\dagger}\phi_2 \\
S_{fr}^{s}  &=& \sum_{i=1}^n \phi_{2+i}^{\dagger}
               [Q(\kappa^s)^{\dagger}Q(\kappa^s)]^{-{1\over 2n}} \phi_{2+i}
\end{eqnarray}
We calculate $S_{fr}$ using the RHMC algorithm~\cite{Clark:2003na} with
optimised values for $n$ and the number of fractions.  Each
term of the action is split into one ultraviolet and two infrared parts,
\begin{eqnarray}
S_{\rm UV} &=& S_{g} \\
S_{\rm IR-1} &=& S_{\Det}^{l} + S_{\Det}^{s} + S_{f1}^{l} \\
S_{\rm IR-2} &=& S_{f2}^{l} + S_{fr}^{s}\,.
\end{eqnarray}
In the leap-frog integrator $S_{\rm UV}$, $S_{\rm IR-1}$ and 
$S_{\rm IR-2}$ are put on {\it three separate} time scales,
\begin{eqnarray}
V(\tau) &=&
\Big[ V_{\rm IR-2} \left({\delta\tau \over 2}\right) ~~A^{m_1}~~ 
      V_{\rm IR-2} \left({\delta\tau \over 2}\right)\Big] ^{n_\tau} \\
A &=& V_{\rm IR-1} \left({\delta\tau \over 2m_1}\right) ~~B^{m_2}~~ 
      V_{\rm IR-1} \left({\delta\tau \over 2m_1}\right) \\
B &=& V_{\rm UV}   \left({\delta\tau \over 2m_1 m_2 }\right)
      V_Q         \left({\delta\tau \over  m_1 m_2 }\right)
      V_{\rm UV}   \left({\delta\tau \over 2m_1 m_2 }\right)
\end{eqnarray}
where $n_\tau =\tau / (\delta \tau)$ and the $V$s are evolution
operators of the Hamiltonian.

\subsection{Solvers}

Besides the standard conjugate gradient \textsl{(cg)} solver BiCGstab
and GMRES were implemented.  Variants with mixed precision arithmetics
are available for \textsl{cg} and BiCGstab.  In order to reduce time
spent in the solver chronological inversion \cite{Brower:1995vx} is
employed and even-odd preconditioning as well as Schwarz preconditioning
\cite{Luscher:2005rx} are used.

%==============================================================================
\section{Implementation details}

The code is mostly written Fortran.  The C preprocessor is used for
preprocessing in general and the \textsl{m4} macro processor for a few files.
A simple mechanism is employed to automatically generate multi precision
versions from the same source.  BQCD is parallelised with MPI and OpenMP.  The
first version of the program was parallelised for a Cray T3E with the
\textsl{shmem} library.  \emph{shmem} can still be used in the hopping matrix
multiplication. 

Random numbers are generated with \textsl{ranlux}
\cite{Luscher:1993dy,ranlux}.  Binary data (SU(3) configurations) can
either be stored in a native BQCD format or in the International Lattice
DataGrid (ILDG) \cite{Beckett:2009cb} format.
The input parameter file and the log file are simple text files that have a
\emph{keyword value(s)} structure.  Important parts of the program are
instrumented for time profiling and performance measurements.

\vfill
%------------------------------------------------------------------------------
\begin{table}[h]
\begin{tabular*}{\textwidth}{cc@{\extracolsep{\fill}}ccc@{\extracolsep{\fill}}ccc}
\hline
        &         & 
\multicolumn{3}{c}{hopping matrix} & 
\multicolumn{3}{c}{\textsl{cg} solver} \\
        &         & 
\multicolumn{3}{c}{multiplication (Fortran)} \\ 
\cline{3-5}\cline{6-8}
        &         & per core& overall & fraction & per core& overall & fraction\\
\#racks & \#cores & Mflop/s & Tflop/s & of peak  & Mflop/s & Tflop/s & of peak \\
\cline{1-2}\cline{3-5}\cline{6-8}
1/2& \D2048   & \D344 & \D0.70 &  10.1\,\%   & \D385 & \D0.79 &  11.3\,\%  \\
 1 & \D4096   & \D429 & \D1.76 &  12.6\,\%   & \D461 & \D1.89 &  13.6\,\%  \\
 2 & \D8192   & \D415 & \D3.40 &  12.2\,\%   & \D444 & \D3.64 &  13.1\,\%  \\
 4 &  16384   & \D407 & \D6.67 &  12.0\,\%   & \D423 & \D6.93 &  12.4\,\%  \\
\hline
\end{tabular*}
\caption{Performance figures for a $48^3\times96$ lattice obtained with
  the pure Fortran implementation on a Blue\,Gene/P.}
\end{table}

%------------------------------------------------------------------------------
\begin{table}[h]
\begin{tabular*}{\textwidth}{cc@{\extracolsep{\fill}}ccc@{\extracolsep{\fill}}ccc}
\hline
        &         & 
\multicolumn{3}{c}{hopping matrix} & 
\multicolumn{3}{c}{\textsl{cg} solver} \\
        &         & 
\multicolumn{3}{c}{multiplication (assembler)} \\ 
\cline{3-5}\cline{6-8}
        &         & per core& overall & fraction & per core& overall & fraction\\
\#racks & \#cores & Mflop/s & Tflop/s & of peak  & Mflop/s & Tflop/s & of peak \\
\cline{1-2}\cline{3-5}\cline{6-8}
1/2& \D2048   &  1057 & \D2.16 &  31.1\,\%   & \D821 & \D1.68 &  24.1\,\%  \\
 1 & \D4096   &  1061 & \D4.35 &  31.2\,\%   & \D802 & \D3.28 &  23.6\,\%  \\
 2 & \D8192   &  1019 & \D8.35 &  30.0\,\%   & \D763 & \D6.25 &  22.5\,\%  \\
 4 &  16384   & \D923 &  15.11 &  27.1\,\%   & \D684 &  11.21 &  20.1\,\%  \\
\hline
\end{tabular*}
\caption{Performance figures for a $48^3\times96$ lattice obtained with
  an assembler implementation of the hopping matrix multiplication on a
  Blue\,Gene/P.}
\end{table}

\pagebreak
%------------------------------------------------------------------------------
\begin{figure}[t]
\epsfig{file=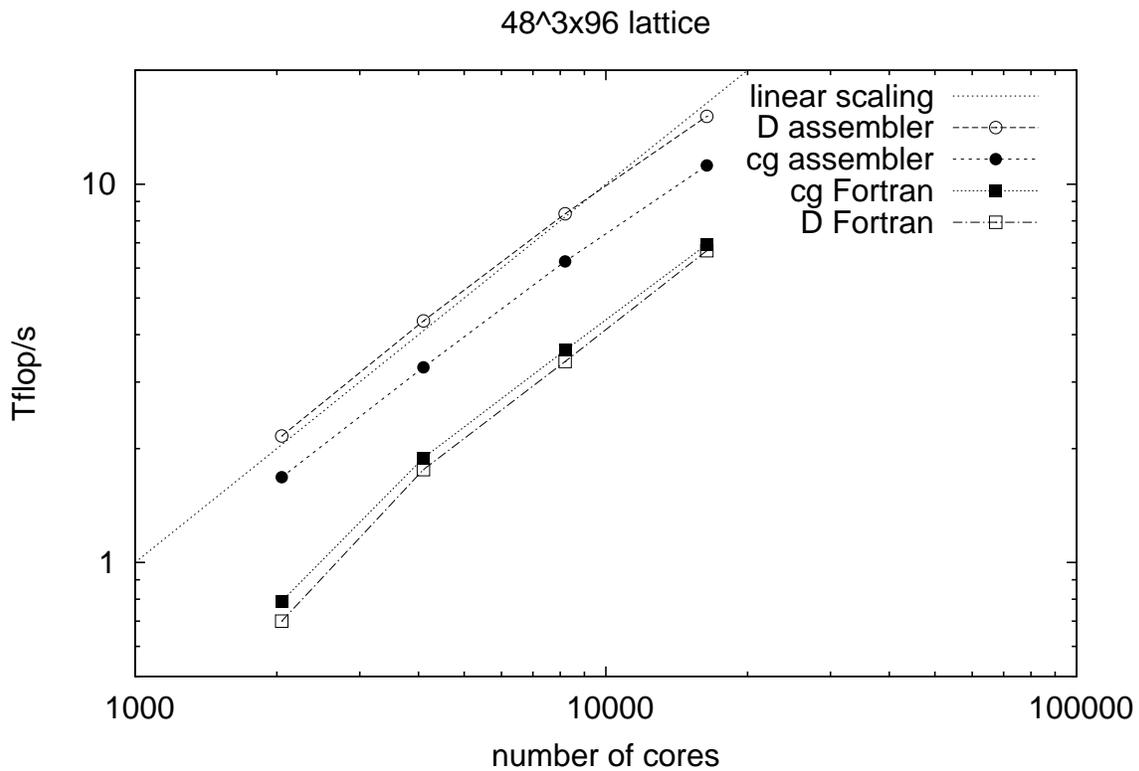,width=\textwidth}
\caption{Scaling plot of performance data from a Blue\,Gene/P given in
  Tables 1 and 2.  The dotted line indicates linear scaling.  Any linear
  scaling runs parallel to this line.}
\end{figure}

%==============================================================================
\section{Performance}

The program scales very well to large numbers of cores.  The pure
Fortran case even displays some super-linear speedup (see Figure~1).  For
Blue\,Gene and Itanium2 assembler implementations of the hopping matrix
multiplication were provided by Th.~Streuer.  Performance figures for a
$48^3\times96$ lattice obtained on a Blue\,Gene/P are given in Tables 1
and 2.  The assembler implementations makes it possible to overlap
communication with computation.  This boosts the performance of the
hopping multiplication of up to a factor of 3.1 and the whole conjugate
gradient solver by a factor of 1.6 to 2.1 compared with the pure
Fortran version.  With this code it is possible to run simulations
at a sustained overall speed of 11.2 Tflop/s 

%==============================================================================
\section{Acknowledgements}

We would like to thank Gerrit Schierholz, Roger Horsley, Dirk Pleiter,
Paul Rakow and James Zanotti for support, stimulating discussions and
bug reports, Thomas Streuer for providing assembler code and Andrea
Nobile for discussions on Schwarz preconditioning.  The computations
were performed on the Blue\,Gene/P at J\"ulich Supercomputer Centre,
J\"ulich, Germany.

\pagebreak
%==============================================================================

\end{document}